\documentclass[aps,superscriptaddress,onecolumn,showpacs,preprintnumbers,amsmath,amssymb]{revtex4}
\usepackage[cp1251]{inputenc}
\usepackage{amssymb,amsmath}
\usepackage[mathscr]{eucal}
\usepackage{graphicx}
\usepackage{longtable}
\usepackage[dvips]{hyperref} 
\begin{document}
\title{Rabi Waves and Peculiarities of Raman Scattering  in   Carbon  Nanotubes, Produced by High Energy Ion Beam Modification of Diamond Single Crystals}
\author{Dmitry Yearchuck (a),  Alla Dovlatova (b)\\
\textit{(a) - Minsk State Higher Aviation College, Uborevich Str., 77, Minsk, 220096, RB; yearchuck@gmail.com,  \\ (b) - M.V.Lomonosov Moscow State University, Moscow, 119899}}
\date{\today}
\begin{abstract}  QED-model for  multichain coupled qubit system, proposed in \cite{Part1}, was confirmed by Raman scattering studies of carbon zigzag-shaped nanotubes, produced by high energy ion beam modification of natural diamond single crystals. New quantum optics phenomenon - Rabi waves - has been experimentally identified for the first time. Raman spectra in perfect quasi-1D carbon  nanotubes are quite different in comparison with well known Raman spectra in 2D carbon  nanotubes of larger diameter. They characterized by vibronic mode of Su-Schriffer-Heeger $\sigma$-polaron lattice and its revival part in frequency representation, which is the consequence of Rabi wave packet formation.  
\end{abstract}
\maketitle 
\section{Introduction and Background}
Quantum electrodynamics (QED) takes on more and more significance for its practical application and it, in fact, becomes to be working instrument in spectroscopy studies and industrial spectroscopy control. 
 QED-model for  multichain coupled qubit system was proposed in \cite{Part1}. It     is generalization of the  model described in \cite{Slepyan_Yerchak}, generalizing, in its turn, Tavis-Cummings model  by taking into account the 1D-coupling between qubits. The most substantial result in \cite{Slepyan_Yerchak} is the prediction of new quantum optics phenomenon - Rabi waves  formation.
  It is substantial, that in the model, proposed in \cite{Part1} the interaction of quantized EM-field with multichain qubit system  is considered by taking into account both the intrachain and interchain qubit coupling. Tne model is presented with an example of perfect zigzag-shaped carbon nanotubes (ZSCNTs). ZSCNTs  can be considered  to be the set of $n$ carbon backbones of trans-polyacetylene (t-PA) chains,  which are connected between themselves. Given  set can be  considered to be a single whole, which  holds the quasionedimensionality of a single chain. It seems to be correct for perfect ZSCNTs, if their diameter is $\leq$ 1nm (see Sec.II). 

 It is substantial, that  the physical properties of  t-PA chains are studied very well both theoretically and experimentally. Let us touch on given subject in more detail.  The authors of \cite{Su_Schrieffer_Heeger_1979}, \cite{Su_Schrieffer_Heeger_1980} have found the most simple way to describe mathematically  the chain of t-PA by considering it to be Fermi liquid. It is now well known SSH-model. The most substantial suggestion in SSH-model is concerned the number of degrees of freedom. Su, Schrieffer, Heeger  suggested, that the only dimerization coordinate $u_n$ of the $n$-th $CH$-group, 
$n = \overline{1,N}$ along chain molecular-symmetry axis $x$ is essential for determination of main physical properties of t-PA. Other five degrees of freedom were not taken into consideration. Nevertheless, the model has obtained magnificent experimental confirmation. In \cite{Part1} was  suggested, that given success is the consequence of some general principle and it was shown, that    given general principle is really exists and main idea was proposed by Slater at the earliest stage of quantum physics era already in 1924. It is - "Any atom may in fact be supposed to communicate with other atoms all the time it is in stationary state, by means of virtual field of radiation, originating from oscillators having the frequencies of possible quantum transitions..." \cite{Slater}. The development of given idea is based on the results of the work \cite {Dovlatova}. It has been found  in \cite {Dovlatova}, that Coulomb field in 1D-systems or 2D-systems can be quantized, that is, it has the character of radiation field and it can exist without the sources, which have created given field. Consequently, Coulomb field can be considered to be "virtual" field  in Slater principle and it can be applied to both t-PA and NTs.  It produces in t-PA the preferential direction in atom  communication the only along chain axis (to be consequence of quasionedimensionality), and it explains qualitatively the success of SSH-model. Since Slater principle can be  equally applied to trans-polyacetylene (t-PA) chains and ZSCNTs, in the case of quasi-1D ZSCNTs it leads to applicability of SSH-model to given NTs, that is to NTs of small diameter despite the strong interaction between the chains.
Experimental confirmation for given conclusion follows
from ESR-studies in rather perfect CZSNTs, produced by
high energy ion modifcation (HEIM) of diamond single
crysatals \cite{Erchak}, \cite{Ertchak}, \cite{Ertchak_Stelmakh}. In the frames of SSH-model  ZSCNT represents itself autonomous dynamical system with discrete circular symmetry consisting of finite number $n\in N$ of carbon backbones of t-PA chains, which are placed periodically along transverse angle coordinate. Longitudinal axes $\{x_i\}, i = \overline{1,n}$, of individual chains can be directed both along element of cylinder and  along generatrix  of any other smooth figure with axial symmetry. It is taken into account, that in the frames of SSH-model,  the adjacent chains, which represent themselves a mirror of each other, will be indistinguishable, since the only one  degree of freedom - dimerization coordinate $u_m$ of the $m$-th $C$-atom, 
$m = \overline{1,N}$ along chain molecular-symmetry axis $x$ is substantial for determination of main physical properties in the frame of SSH-model. 

The aim of given letter is to represent the experimental evidence for multichain QED-model, proposed in \cite{Part1} and to confirm the theoretical prediction of  Rabi wave phenomenon in \cite{Slepyan_Yerchak}. The new phenomenon predicted is quantum coherent effect. It means, that along with requirement of quasionedimensionality the  requirement of structural perfectness arises. To observe quantum optical coherent effects on NTs, the ensemble of NTs has to be homogeneous. It means, that any dispersion in axis direction, chirality, length and especially in diameter both for single NT along its axis and between different NTs in ensemble has to be absent, axial symmetry has to be also retained, that is, there are additional requirements in comparison with, for example, t-PA technology. The CVD-technology of NTs production and many similar to its ones seem to be not satisfying to above-listed requirements at present. It means, that experimental results and their theoretical treatment will be different in both the cases, that really takes place (see for more details Sec.II). The situation seems to be analogous to some extent to the solid state physics of the same substance in single crystal and amorphous forms. The technology, based on HEIM,  satisfy given requirements.  
Let us give short review concerning HEIM-method of production of incorporated carbon NTs. The first report on the discovery of new carbon phase - carbon nanotubes, incorporated in diamond matrix is related to 1990, and it  was made during 1990 IBMM-Conference, Knoxwille, USA and it was repeated at E-MRS 1990 Fall Meeting, Strasbourg, France, \cite{Efimov}, that is, nothing was at that time known on the japan discovery of free NTs, related to 1991.
 It was found, that central place takes in ESR spectra of diamond single crystals, modified by high energy ion implantation, the strong line with very unusual radiospectroskopic properties. They was anisotropic, however the anisotropy was  weak in comparison with the anisotropy of point centers in diamond. Especially interesting, that along with $g$-value  the linewidth $\Delta{H_{pp}}$ was also found to be tensor quantitity, at that $g$-tensor and $\Delta{H_{pp}}$-tensor were characterized by the same symmetry group\cite{Erchak}. Moreover the only one eqivalent configuration was presented with principal direction along ion beam direction for both $g$ and $\Delta{H_{pp}}$ angular dependencies.  Other lattice equivalent configurations were absent. The kind of symmetry group, corresponding to  $g$-tensor and $\Delta{H_{pp}}$-tensor symmetry was strongly dependent on  the choice of ion beam direction relatively the crystallographic axes of diamond lattice.  It was found for the first time in radiospectroscopy, that $g$-tensor and $\Delta{H_{pp}}$-tensor symmetry are the mapping of the symmetry of lengthy objects with macrosizes along implantation direction \cite{Erchak}. Detailed studies of ESR spectral  angular dependencies allowed to establish, that the  structures formed by $\langle 111 \rangle$-HEIM in diamond have tracklike cylindrical shape, which are lengthy strongly in $\langle 111 \rangle$-direction, with the size varying from several mkm to several tens of mkm, depending on ion energy. In other words, the first studies in 1990 allowed to establish the formation of carbon nanotubes of cylindical symmetry shape incorporated in diamond matrix. Simultaneously the formation of quite different nanotubes was established. They represent themselves crimped  cylinder with crimping corresponding to four-petal structure in the cross section.  They are produced by the implantation direction, coinciding with $\langle 100 \rangle$ axis of diamond lattice \cite{Erchak}.  The next step was the identification of the structure of cylindical symmetry shape $\langle 111 \rangle$-incorporated nanotubes with rolled up graphene sheet in zigzag shaped configuration. It was done by the study of radiospectroscopic properties of spin curriers in given tubes. It was found, that the system of paramagnetic centers (PC), which are responsible for appearance of strong absorption above described is non-Blochian system. A number of distinctive peculiarities have been observed for the first time in radiospectroscopy at all. The main ones among them are the following.

1.It was found, that the shape of resonance lines cannot be described by Lorentzian, Gaussian, or by their convolution. The shape is characterized by essentially more slow decrease of the absorption intensity on the wings in comparison with known shape functions. It was called super-Lorentzian. It was proved, that super-Lorentzian is genuine lineshape of resonance absorption (that is, it does not represent the superposition of  Lorentzians with different linewidths) \cite{Ertchak}.

2.It was established, that PC-system is nonsaturating with superlinear absorption kinetics (that is, the dependence of resonance signal amplitude on the amplitude of  magnetic component of microwave field is  superlinear). Especially interesting, that superlinear absorption kinetics  has been observed by both the registration of resonance signal in phase with high-frequency modulation  and in  quadrature  with high-frequency modulation \cite{Erchak}, \cite{Ertchak}.

3.It was found unusual for nonsaturating resonance systems dependence of resonance signal amplitude on the value of modulation frequency, which was characterized by substantial increase (instead decrease) of resonance signal amplitude with modulation frequency increasing \cite{Erchak}, \cite{Ertchak}.

4.It was established the effect of  appearance  of "phase angle", characterizing the absorption process. It consist in that, that despite nonsaturating absorption kinetics, the maximum of absorption is achieved not strictly in phase with modulation field, that is at modulation phase value, which is nonequal to zero. "Phase angle" is anisotropic in general case and maximum of its value was found to be equal $20\textdegree$ \cite{Ertchak}. 

The analysis of given peculiarities \cite{Ertchak}, \cite{Ertchak_Stelmakh} has lead to  conclusion, that PC, which are responsible for strong non-Blochian absorption, are mobile and are characterized  by two times of spin-lattice relaxation, by $T_1$, which is comparable with relaxation time  of usual point PC in diamond single crystals $( 10^{-3} - 10^{-4} s)$ and by very short time $\tau$ (upto $10^{-13} s$)  of conversion of the energy of the spin system into mechanical kinetic energy 
of $PC$ motion, which can be considered to be consequence of fundamental in magnetic resonance phenomenon of gyromagnetic coupling between magnetic and mechanical moments. Therefore $\tau$-process is very fast process in comparison with any process, realized by means of pure  magnetic interaction, the  relaxation times of which  are not shorter, than $10^(-10) s$ \cite{Abragam}.  
The possibility of given conversion  can be realized naturally, if the energy of the motion activation is very small, that  is typical for mobile topological solitons like them identified in $t$-PA. So the conclusion, that the strong non-Blochian absorption is determined by mobile topological solitons was obtained by natural way. All subsecuent studies have confirmed given  conclusion. It was found, that there is numerical coincidence of the characteristics of SSH topological $\pi$-solitons in t-PA and in  $\langle 111 \rangle$ incorporated NTs. So, the values of g-tensor components in $Cu$-implanted sample are $g_1$ = 2.00255 (it is minimal $g$-value and it is $g_\mid\mid$ principal direction of axial $g$-tensor, at that it coincides with ion beam direction), $g_2$ = $g_3$ = $g_\perp$ = 2.00273, the accuracy of relative g-value measurements is $\pm 0.00002$ \cite{Erchak}. $g$-value of paramagnetic $\pi$-solitons in trans-polyacetylene, equaled to 2.00263 \cite{Goldberg}, gets in the middle of given rather narrow interval of $g$-value variation of PC in ion produced NT. Although anisotropy of paramagnetic $\pi$-solitons in $t$-PA, which are also considered to be mobile paramagnetic centers, mapping the distribution of $\pi$-electron density along whole individual $t$-PA chain, is not resolved by ESR measurements directly (which in fact is the indication, that chemically
produced $t$-PA is less perfect in comparison with NTs in diamond matrix), there are indirect evidences on axial symmetry of ESR absorption spectra in $t$-PA too, \cite{Kahol}, \cite{Kuroda}. Consequently, the value 2.00263 is mean value and it coincides with accuracy 0.00002 with mean value of aforecited principal $g$-tensor values of PC in NTs. Given coincidence becomes to be understandable now, if to take into account the results of \cite{Part1}, where is shown, that SSH-model is applicable to zigzag NTs, produced by HEIM of diamond single crystals. 

Thus we see, that it was possible to determine anisotropy of $g$-values with very high precision, that indicates on the very perfect and homogeneous NTs, produced with strict axial symmetry along ion beam direction and to establish the origin of paramagnetic centers  in ion produced NT to be paramagnetic topological $\pi$-solitons (SSH-solitons), considered to be mobile PC, mapping the distribution of$ \pi$-
electron density along whole individual NT \cite{Ertchak}, \cite{Ertchak_Stelmakh}.  
   Thereby immediately from ESR studies was found, that structural element of new carbon phase with cylindrical symmetry produced in diamond by $\langle{111}\rangle$ high energy implantationis is t-PA chain backbone.  At the same time t-PA chain backbone is basic element for both  graphene, graphite, NTs formation and also for the carbyne formation. Carbynes were studied by ESR, they characterized by quite different ESR-spectra \cite{Ertchak_J_Physics_Condensed_Matter}.  The possibility of formation of cylindrical graphite rod  is also not corresponding, since  graphite is characterized by other ESR spectra. So it was concluded unambiguously, that the spectra observed by $\langle{111}\rangle$ high energy implantation are corresponding the only to graphene sheet, which is rolled up in that way, in order to t-PA chain backbones were directed along $\langle{111}\rangle$-axis of diamond lattice, coinciding with ion beam direction resulting in CZSNTs production. 

It is substantial, that free nanotubes were considered theoretically   to be $2D$-strutures and two dimensional lattice structure of a single wall carbon nanotube (SWNT)
is specified uniquely by the chirality defined by two integers
$(n,m)$ \cite{Dresselhaus}, \cite{Saito}. In the Raman spectra of a SWNT are  two in-plane $G$ point longitudinal
and transverse optical phonon $(LO and TO)$ modes \cite{Reich} and
the out-of-plane radial breathing mode $(RBM)$ \cite{Dresselhaus M.S} are observed.
The $LO$ and $TO$ phonon modes at
the $G$ point in the two-dimensional Brillouin zone
are degenerate in graphite and graphene, however thy split in SWNT into two peaks,
denoted by $G^+$ and $G^-$peaks, respectively, \cite{G.Dresselhaus} because of the
curvature effect. The agreement of experimental Raman studies of NTs with diameter $ > 1 nm$ with $2D$ SWNT-theory unambiguously indicates, that given NTs, produced  by CVD are really $2D$ systems.  At the same time the narrow perfect tubes with diameter $ < 1 nm$ cannot be considered strongly speaking to be  $2D$-systems, they are quasi-$1D$ systems. Really, in \cite{Wang} is reported on the development of  SWCNTs of 0.4 $nm$ diameter – the smallest so
far - inside the nanochannels of porous zeolite $AlPO4-5$  single crystals and the authors are considered 0.4 $nm$ diameter NTs to
be ideal one-dimensional quantum hollow wires. The electronic and magnetic properties
of these ultra-small NTs seem to be drastically different from those of large sized
NTs.  Really given tubes were shown to exhibit unusual
novel phenomena like diamagnetism and superconductivity at low temperatures \cite{Tang} in
addition to several other optical properties and doping induced effects.
It seems to be the consequence of onedimensionality of given NTs.
The direct indication to given conclusion is recent ESR-studies in \cite{Rao}, where   
electron spin resonance
measurements on ultra-small single walled carbon nanotubes embedded in a SAPO 5
zeolite matrix with a main point of attention to potentially occurring CESR (ESR on the electrons in c-band) signals.
Instead, only one paramagnetic signal is observed of symmetric shape at $g = 2.0025$ on the CNTs in zeolite cages, the ESR
signal exhibits a predominantly Gaussian character at $8.9 GHz (4.2 K)$, with a $\Delta{B_{pp}} = 8.3 G$ and $g = 2.00251$. For the free standing SWCNTs, the ESR signal takes a more
Lorentzian shape at $8.9 GHz (4.2 K)$, now with $\Delta{B_{pp}} = 5 G$ and with $g = 2.00254$. So,
from the above observed distinct changes in the ESR spectral properties of NTs with
and without cages, it was inferred in \cite{Rao}, that the ESR signal indeed stems from the carbon NTs
presenting inside the nanochannels of the zeolite. The  observed $g$-value (2.00251) is
inconsistent with the earlier measured values $(g = 2.05 - 2.07)$ \cite{Chauvet}  of CESR on non
embedded single- and multiwalled CNTs of larger diameter.
There was observed in \cite{Rao} a dependence of peak-to-peak ESR signal
width in confined SWCNT@SAPO 5 at 70 $K$  on microwave frequecy $f$
suggesting a  inhomogeneous (Gaussian) contribution to the line broadening.
 Least–square linear fit of $\Delta{B_{PP}}$  is $B_{PP}(G)$ = $7.7 + 0.018 f (GHz)$. It is clear evidence, that the so called ultra-small carbon nanotubes, studied in \cite {Rao} are not perfect. Nevetheless they can be really considered to be quasionedimensional objects. It means, that like to $t$-PA,  Peierls transition has to take place. Direct indication to given conclusion is $g$-value, which corresponds the only to deep centers in bandgap in any carbon materials \cite{Ertchak}. On the one hand it is indicating of the appearance of bandgap in starting metal tubes to be consequence of Peierls metal-semiconductor transition. On the other hand $g = 2.00254$ is  coincides {in limits of accuracy of measurements) with $g$-value of  $g_1$ = 2.00255 of axial $g$-tensor of  SSH $\pi$- -solitons in CZSNT, incorporated in diamond matrix  by $\langle 111 \rangle$ diamond HEIM and it near to the $g$-value of paramagnetic SSH $\pi$-solitons
in trans-polyacetylene, equaled to 2.00263 \cite{Goldberg}, that indicates on the similarity of the structure of tubes incorporated in diamond and embedded in zeolite. At the same time angular dependences of $g$-value and $\Delta{B_{pp}}$ were not observed  in \cite {Rao}. It seems to be additional argument to conclusion on insufficient homogeneity of NTs in cited work. Really they have different chiralities (authors report on  the presence of  three chiralities (5,0) (4,2) and (3,3)). On the other hand the characteristic for spin-glass temperature dependence of $\Delta{B_{pp}}$ seems to be the indication, that the  SSH $\pi$ -solitons are pinned and have random  distribution along tube axis. For comparison, the temperature dependence of $\Delta{B_{pp}}$ in NTs, produced by HEIM is quite different and it indicates, that   $\Delta{H_{pp}}$ is not determined by unresolved g-anisotropy or unresolved hyperfine interaction \cite{Erchak}. It is interesting, that pinned solitons can be considered to be similar to chemical radicals, possessing by dangling bonds, which to some extent is in agreement with interpretation of PC-nature to be dangling bond defects  in \cite{Rao}, however they are not usual dangling bond point defects, but they are topological defects. Random  distribution along tube axis means in turn, that individual tube withdefined chirality is also not sufficiently perfect to observe optical quantum coherent effects. It means, that Slater principle  with one preferential direction will be not applicable and optical spectra, in particular Raman spectra, will be determined the only by local properties of optical centers, that is, they can be similar to those ones in the tubes of larger diameters. It is substantial, that main Raman modes in the tubes of larger diameters seem to be similar in both perfect and in not sufficiently perfect NTs, since even in perfect NTs in given case two preferential direction are existing and onedimensional quantum coherent SSH-model \cite{Part1} will be not applicable for description of optical properties in given tubes.

 \section{Results and Discussion}

Samples of type IIa natural diamond, implanted by high energy ions of copper $(63$ $MeV$, $5\times{10^{14}}$ $cm^{-2}$) and boron $(13,6$ $MeV)$ have been studied. Ion implantation was performed along $\left\langle{111}\right\rangle$ crystal direction. Raman scattering  (RS) spectra were registered in backscattering geometry. Laser excitation wave length was 488 $nm$, rectangular slit $350{\times}350 (\mu m)^2$ was used, scan velocity was 100 $cm^{-1}$ pro minute.
 
The spectra observed are presented in Figures 1 to 3. 
All the spectra  are $180\textdegree$ out of phase, that is dip positions correspond to maxima of the signals.   The spectrum, presented in Figure 1,  is characteristic of the only ion beam modified region (IBMR) of the sample, since the RS-line near 1332 $cm^{-1}$, characteristic for diamond single crystals, is absent in the spectrum. The RS-lines with peak positions $656.8{\pm}0.2$ $cm^{-1}$, $1215{\pm}1$ $cm^{-1}$, $1779.5{\pm}1$ $cm^{-1}$ and  $2022.3{\pm}0.5$ $cm^{-1}$ correspond to optical centers in IBMR   by laser excitation transversely to sample surface from implanted side. At the same time, by laser excitation of the same sample from opposite unimplanted side the qualitatively other picture is observed, see Figure 2. Firstly, the line,  characteristic for diamond single crystals, is now  presented in the spectrum (the whole amplitude of given line is not shown in Figure 2). It is interesting, that its frequency value (1328.7 $cm^{-1}$) is slightly shifted from usually observed value near 1332 $cm^{-1}$. It  indicates on renormalization of optical phonon by Cu-implantation in the whole diamond matrix.

\begin{figure}
\includegraphics[width=0.5\textwidth]{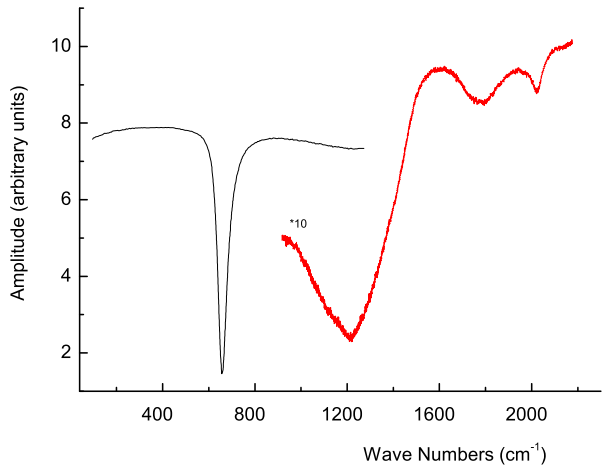}
\caption[Spectral distribution of Raman scattering  intensity in
diamond single crystal, implanted by high energy copper ions, the excitation is from implanted side of the sample.]
{\label{Figure1} Spectral distribution of Raman scattering  intensity in
diamond single crystal, implanted by high energy copper ions, the excitation is from implanted side of the sample.}
\end{figure}
 \begin{figure}
\includegraphics[width=0.5\textwidth]{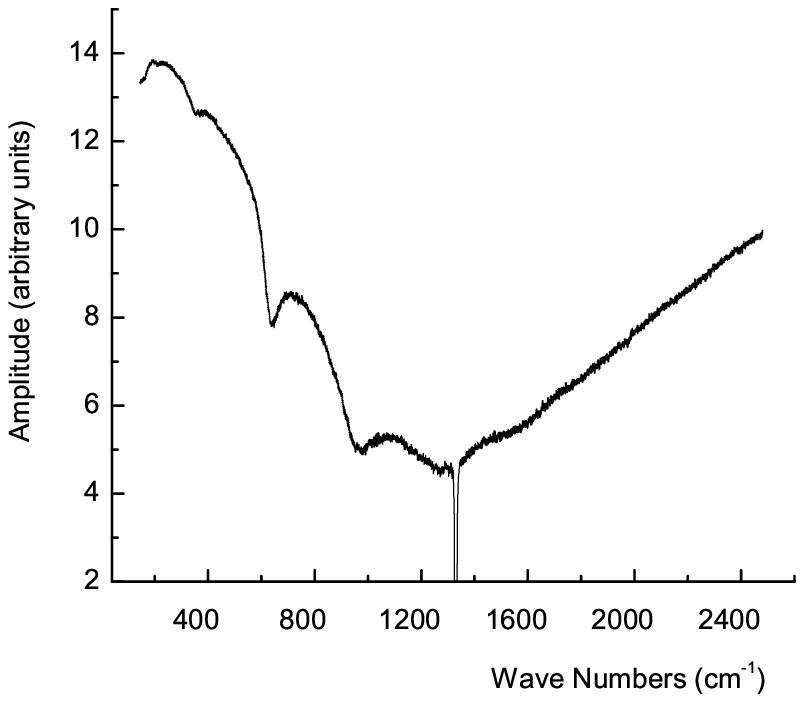}
\caption[Spectral distribution of Raman scattering  intensity in
diamond single crystal, implanted by high energy copper ions, the excitation is from unimplanted side of the sample.]
{\label{Figure2} Spectral distribution of Raman scattering  intensity in
diamond single crystal, implanted by high energy copper ions, the excitation is from unimplanted side of the sample.}
\end{figure}
\begin{figure}
\includegraphics[width=0.5\textwidth]{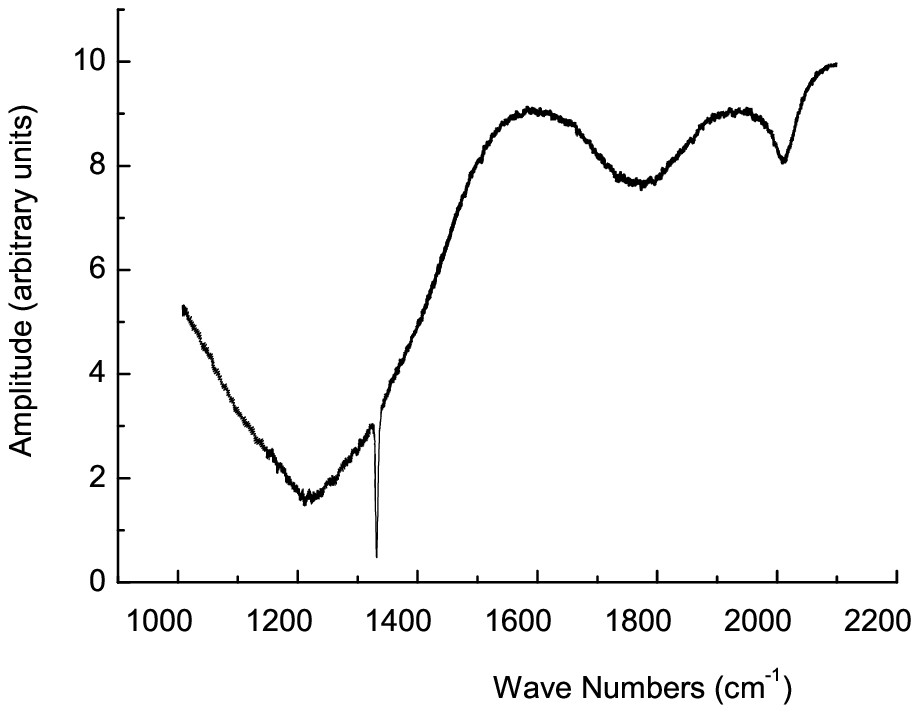}
\caption[Spectral distribution of Raman scattering  intensity in
diamond single crystal, implanted by high energy boron ions, the excitation is from implanted side of the sample.]
{\label{Figure3} Spectral distribution of Raman scattering  intensity in
diamond single crystal, implanted by high energy boron ions, the excitation is from implanted side of the sample.}
\end{figure}
    
 Secondly, RS-lines with peak positions  at 354.6, 641.8, 977.1 (${\pm}1$ $cm^{-1}$), 1274.1 ${\pm}2$ $cm^{-1}$  and more weak pronounced lines at 1569 ${\pm}3$ $cm^{-1}$, 1757${\pm}5$ $cm^{-1}$ were observed. Moreover, all the lines listed are superimposed now with very broad (its linewidth  is 1720${\pm}20$ $cm^{-1}$) asymmetric line with peak position $1160{\pm}10$$cm^{-1}$. 
The spectrum of boron implanted sample, Figure 3, was measured in the range (1000 - 2100) $cm^{-1}$. It is seen, that the spectra of boron and copper implanted samples are qualitatively similar in given spectral range. However numerical values  - 1212.3${\pm}1$,  1772.5${\pm}1$, 2011${\pm}0.5$$cm^{-1}$ - of the peak positions are slightly different, they are shifted to low frequency region. It is interesting, that there is regularity in peak position shift, which is increasing  with frequency increase, and it is equal to 2.8, 7, 11.3 $cm^{-1}$ correspondingly.  It is strong indication, that all three lines belong to the same optical system in both the samples studied. The  line with peak position 1331.95 ${\pm} 0.1$$cm^{-1}$ was also presenting. Its frequency value coincides with the value of optical phonon peak in conventional natural diamonds. The presence of given RS-line by excitation from implanted side can be determined by two factors. Firstly, the effective thickness of IBMR is substantially less, it is $\simeq{1.4}\mu m$, in comparison with Cu-implanted sample ($\simeq{7}\mu m$) \cite{Erchak}, consequently, laser excitation can reach an unimplanted region. It can also be suggested, that the modification is not entire in near surface region.  Given results seem to be the first results in application of RS-spectroscopy to study of IBMR in diamonds.  

Comparing RS-spectra, represented in figures 1 to 3 with well known spectra of NTs produced by CVD-methods it is seen, that difference between the optical characteristics of the NTs, produced by HEIM of diamond single crystals and  between the NTs, produced by other methods, including ultra-small NTs, is  greatly despite the essential similarity of ESR-data with those ones for ultra-small NTs, embedded in zeolite matrix.   On the other hand,
comparing the figures 1 and 2  between themselves, we can straight away conclude, that RS-lines observed cannot be attributed to usual atomic or molecular vibrations including the local vibration modes like to well known G-bands and  radial breathing modes, observed in  NTs, produced by CVD-methods, since the RS-spectra, corresponding to usual local vibrations cannot be dependent on the change of  propagation direction of exciting light into opposite. Therefore 2D theory, developed for explanation of optical properties of NTs cannot be applicable for perfect NTs of small diameter. It is strong argument for  application of the QED-model, proposed in \cite {Part1} for correct description of  the objects studied.
Since the theoretical  model of small 1D NTs \cite {Part1} is based on SSH-model of 1D organic conjugated conductors let us consider SSH-model more carefully. It will be shown, that it needs in some corrections and development. 

We have repeated the calculation, presented  in \cite{Su_Schrieffer_Heeger_1979}, \cite{Su_Schrieffer_Heeger_1980}. It has been found, that there is the second solution, which leads to independent branch of quasiparticles. Let us show it. We preserve all designation for quantities from \cite{Su_Schrieffer_Heeger_1980}. We have obtained the expression for Hamiltonian in reduced zone in the form
\begin{equation}
\label{Eq1}
\hat{H}(u) = \sum_{k}\sum_{s}[\varepsilon_k ({\alpha}^2_k - {\beta}^2_k) + 2 {\alpha}_k{\beta}_k \Delta_k](\hat{n}^c_{ks} - \hat{n}^v_{ks}) + 2NKu^2\hat{e},
\end{equation}
where ${\alpha}_k$ and ${\beta}_k$  are determined by
\begin{equation}
\label{Eq2}
\alpha_k = \sqrt {\frac{1 \mp \frac{\varepsilon_k}{E_k}}{2}},
{\beta}_k = \sqrt {\frac{1 \pm \frac{\varepsilon_k}{E_k}}{2}}.
\end{equation}
$E_k$ is determined by expression
\begin{equation}
\label{Eq3}
E_k = \sqrt{\varepsilon^2_k  + \Delta^2_k},
\end{equation}
in which
\begin{equation}
\label{Eq4}
\varepsilon_k = 2t_0\cos{ka},
\Delta_k = 4\alpha u \sin{ka}
\end{equation}
and $\hat{n}^c_{ks}$, $\hat{n}^v_{ks}$ are occupation number operators in $C$- and $V$-bands correspondingly.
The known SSH-solution corresponds to choice of lower signs in relations (\ref{Eq2}). The choice of upper signs gives independent solution. Therefore, for the energy of quasiparticles $E^{c[u]}_k$, $E^{v[u]}_k$ in $C$- and $V$-bands, corresponding to upper-sign-solution we obtain
\begin{equation}
\label{Eq5}
E^{c[u]}_k = \frac{\delta E^{[u]}(u)}{\delta{n}^c_{ks}} = \frac{\Delta^2_k - \varepsilon^2_k}{E_k}, E^{v[u]}_k = \frac{\delta E^{[u]}(u)}{\delta{n}^v_{ks}} =  -(\frac{\Delta^2_k - \varepsilon^2_k}{E_k}), 
\end{equation}
We see, that it differs from the energy of  quasiparticles $E^{c[l]}_k$, $E^{v[l]}_k$ in $C$- and $V$-bands, corresponding to lower-sign-solution (SSH-solution), which is 
\begin{equation}
\label{Eq6}
E^{c[l]}_k = \frac{\delta E^{[l]}(u)}{\delta{n}^c_{ks}} = E_k, E^{v[l]}_k = \frac{\delta E^{[l]}(u)}{\delta{n}^v_{ks}} = -E_k.
\end{equation} 
$E^{[u]}(u)$ and  $E^{[u]}(u)$ in (\ref{Eq5}) and (\ref{Eq6}) are eigenvalues of operator $\hat{H}(u)$, which correspond to upper and lower signs in (\ref{Eq2}), 
 ${n}^c_{ks}$, ${n}^v_{ks}$ are eigenvalues of operators of particle numbers in $C$-band and $V$-band correspondingly. In \cite{Su_Schrieffer_Heeger_1980} the coefficients  ${\alpha}_k$ and ${\beta}_k$  were found from the conditions of minimum for the energy. However the only necessary conditions were used. At the same time the  sufficient conditions for the minimum are substantial in given case, they change the role of both solutions. Really, since  ${\alpha}_k$ and ${\beta}_k$ are coupled by the condition
\begin{equation}
\label{Eq7}
{\alpha}^2_k + {\beta}^2_k = 1,
\end{equation} 
 we have conditional extremum and  sufficient conditions for the minimum
 are obtained by standard study.  The second differential of  the energy to be the function of  three variables  ${\alpha}_k$,  ${\beta}_k$ and $\lambda$, where 
 $\lambda$ is coefficient in Lagrange function for conditional extremum, has to be positively defined quadratic form. From the condition of positiveness of three principal minors of quadratic form coefficients we obtain  the following three sufficient conditions for the energy minimum

\paragraph{The first condition}

\begin{equation}
\label{Eq8}
\{ \varepsilon_k (1 - \frac{\varepsilon_k}{E_k}) < \frac{\Delta^2_k}{E_k} | ({n}^c_{ks} - {n}^v_{ks}) < 0\}, \{\varepsilon_k (1 - \frac{\varepsilon_k}{E_k} > \frac{\Delta^2_k}{E_k} | ({n}^c_{ks} - {n}^v_{ks}) > 0 \}
\end{equation}
for the SSH-solution and

\begin{equation}
\label{Eq9}
\{ \varepsilon_k (1 + \frac{\varepsilon_k}{E_k} < \frac{\Delta^2_k}{E_k} | ({n}^c_{ks} - {n}^v_{ks}) < 0 \}, \{\varepsilon_k (1 + \frac{\varepsilon_k}{E_k} > \frac{\Delta^2_k}{E_k} | ({n}^c_{ks} - {n}^v_{ks}) > 0 \}
\end{equation}
 for the additional solution. It is seen, that the first condition is realizable for the quasiparticles of both the kinds, at that in equilibrium $({n}^c_{ks} - {n}^v_{ks} < 0)$  and in  nonequilibrium $(n^c_{ks} - {n}^v_{ks} > 0$ conditions.  
\paragraph{The second condition}
The second condition is the same for both the solutions and it is
\begin{equation}
\label{Eq10}
 (\frac{\varepsilon^2_k}{E_k} - 2\frac{\Delta^2_k}{E_k})^2 - E^2_k  + \frac{3}{4} \Delta^2_k > 0 
\end{equation}

\paragraph{The third condition}
For the SSH-solution we have
\begin{equation}\label{Eq11}
(3\frac{\Delta^2_k}{E_k} + 4\frac{\varepsilon^2_k}{E_k})({n}^c_{ks} - {n}^v_{ks}) > 0. 
\end{equation}

It  means,  that   SSH-solution is unapplicable for description of standard processes, passing near equilibrium state by any parameters. The quasiparticles, described by   SSH-solution, can be created the only in strongly nonequilibrium state with inverse                                                                                                   
population of the levels in $C$- and $V$-bands. At the same time for the solution, which corresponds to upper signs in (\ref{Eq2}), we obtain 
\begin{equation}
\label{Eq12}
(3\frac{\Delta^2_k}{E_k} - 4\frac{\varepsilon^2_k}{E_k})({n}^c_{ks} - {n}^v_{ks}) > 0, 
\end{equation}
which is realizable both in near equilibrium and in strongly nonequilibrium states of the systems by corresponding choose of parameters.

Let us consider the continuum limit for the ground state of the $t$-PA chain with quasiparticles of given branch. Taking into account, that in ground state ${n}^c_{ks} = 0$, ${n}^v_{ks} = 1$ we have
\begin{equation}
\label{Eq13}
E^{[u]}_0(u) = - \frac{2Na}{\pi}\int\limits_0^{\frac{\pi}{2a}} \frac{\Delta^2_k - 
\varepsilon^2_k}{ \sqrt{\Delta^2_k + 
\varepsilon^2_k}}dk + 2NKu^2,
\end{equation}
then, calculating the integral, we obtain
\begin{equation}
\label{Eq14}
E^{[u]}_0(u) =  \frac{4Nt_0}{\pi}\{F(\frac{\pi}{2}, 1 - z^2) + \frac{1 + z^2}{1 - z^2}[E(\frac{\pi}{2}, 1 - z^2) - F(\frac{\pi}{2}, 1 - z^2)]\} + 2NKu^2, 
\end{equation}
where
 $F(\frac{\pi}{2}, 1 - z^2)$ is the complete elliptic integral of the first kind, 
$E(\frac{\pi}{2}, 1 - z^2)$ is the complete elliptic integral of the second kind,
$z^2 = \frac{2\alpha u}{t_0}$.
Approximation of ({\ref{Eq14}}) at $z \ll 1$ gives
\begin{equation}
\label{Eq15}
E^{[u]}_0(u) = N \{\frac{4t_0}{\pi} - \frac{6}{\pi}\ln\frac{2t_0}{\alpha u} \frac{4 \alpha^2 u^2}{t_0} + \frac{28 \alpha^2 u^2}{\pi t_0} + ...\} + 2NKu^2.
\end{equation}
It is seen from (\ref{Eq15}), that energy of quasiparticles, described by   solution, which corresponds to upper signs in (\ref{Eq2}) has the form of Coleman-Weinberg potential with two minima at the values of dimerization coordinate $u_0$ and $-u_0$ like to energy of quasiparticles, described by   SSH-solution \cite{Su_Schrieffer_Heeger_1980}.

It is remarkable, that SSH-model contains in implicit form along with the physical basis for the existence  of solitons, polarons, breathers, formed in $\pi$-electronic subsystem, also the basis for the existence of similar quasiparticles in  $\sigma$-electronic subsystem. The origin  is the same two-fold degeneration of ground state  of the whole electronic system, energy of which in ground state has the form of Coleman-Weinberg potential with two minima (\ref{Eq15}) at the values of dimerization coordinate $u_0$ and $-u_0$. Really the appearance of $u_0 \neq 0$ and $-u_0 \neq 0$ is concerned to the change of interatomic distance. It means, that simultaneously with $\pi$- subsystem, $\sigma$-subsystem is also dimerized. 

  SSH-$\sigma$-polarons have recently been  found in  carbynes \cite{Yearchuck_PL}, and the formation of $\sigma$-polaron lattice (PL), which is antiferroelectrically ordered,  was offered.  Given chain state is  characterized by  the set of lines in IR-spectra, which were assigned with  antiferroelectric spin wave resonance (AFESWR).
Central mode is  antiferroelectric  resonance (AFR)  mode, its value $\nu^\sigma_p(C)$ in carbyne sample was equal to 477 $cm^{-1}$, the splitting parameter in IR detected AFESWR spectra was equal to 150 $cm^{-1}$. 
Known values of $ \nu^\sigma_p(C)$ and AFESWR-splitting parameter   in carbynes allow to estimate the range for expected values of analogous parameters in  CZSNTs. AFESWR-splitting parameter is determined by exchange integrals in $\sigma$-electronic subsystem \cite{Yearchuck_PL}, which seems to be the same in carbynes and in  CZSNTs, since the role of quite different $\pi$-subsystems in carbynes and CZSNTs can be neglected to a first approximation. Consequently, the value of AFESWR-splitting parameter in  CZSNTs (and in t-PA) has to be close to 150 and to 300 $cm^{-1}$ by IR- and RS-AFESWR-detection correspondingly \cite{Yearchuck_Doklady}. The frequencies $\nu^\sigma_p(C)$ and $\nu^\sigma_p(NT)$ of main AFESWR-mode in SSH-$\sigma$-polaron lattice in carbyne and in  CZSNTs
 depend on intracrystalline field \cite{Yearchuck_PL}, that means, that their values will be different. However we can evaluate $\nu^\sigma_p(NT)$, if to take into account  the known relation for the vibration frequencies of similar centers,
the fact of $\sigma$-polaron and $\pi$-soliton lattice formation in carbynes, leading to change in effective masses \cite{Yearchuck_ArXiv}, presence of two  $\pi$-subsystems  in carbynes,  difference of coherence lengths  of $\sigma$-solitons ($\xi_\sigma$) and $\pi$-solitons ($\xi_\pi$), band structure  data  for t-PA  \cite{Grant} and carbynes \cite{Leleiter}.  Relations for coherence lengths \cite{Lifshitz} are 
\begin{equation}
\label{Eq1a}
\xi_{0\pi} = \frac{\hbar v_F}{\Delta_{0\pi}}, \xi_{0\sigma} = \frac{\hbar v_F}{\Delta_{0\sigma}},
\end{equation}
where $\Delta_{0\sigma}$, $\Delta_{0\pi}$ are $\sigma-$ and $\pi$-bandgap values at $T = 0 K$, $v_F$ is Fermi velocity.
We obtain the frequency range for IR SSH-$\sigma$-polaron lines  in t-PA and in CZSNTs
$\nu^\sigma_p(t-PA) \in$ (386.7, 603) $cm^{-1}$ and $\nu^\sigma_p(NT) \in $ (402.5, 627.6) $cm^{-1}$.  Known IR-mode with the frequency near 540 $cm^{-1}$, in t-PA \cite{Heeger_1988} gets to interval (386.7, 603) $cm^{-1}$ and can represent itself AFR  mode in $\sigma$-polaron lattice, that is, there is alternative interpretation of given IR-mode, ascribing earlier to Goldstone SSH-$\pi$-soliton vibration mode \cite{Heeger_1988}. 
From results of \cite{Eklund} follows, that the same  spectral interval with slightly different right-hand value, equaled to  673,7 $cm^{-1}$, is the evaluation for the frequency of $\sigma$-polaron main AFESWR-mode   in  CZSNTs, which is RS-active.  The calculation  in \cite{Eklund} does not take into consideration the soliton and polaron formation. However $\sigma$-polaron  formation does not violate the symmetry of task, that allows to conclude, that the same asymmetry value will be retained  in IR  and RS spectral distributions. 
 Therefore, we come to  conclusion, that intensive line $656.8{\pm}0.2$ $cm^{-1}$ in RS-spectrum, presented in Fig.1, can be assigned with AFR mode of $\sigma$-polaron lattice, produced in NTs. The lines, $1215{\pm}1$ $cm^{-1}$, $1779.5{\pm}1$ $cm^{-1}$ and  $2022.3{\pm}0.5$ $cm^{-1}$ is revival part \cite{Slepyan_Yerchak} of Rabi wave packet in its frequency representation \cite{Part1}. The arguments are the following.
QED-Hamiltonian, proposed in \cite{Part1} for $\sigma$-polaron system of zigzag $NT$, is hypercomplex operator n-number
\begin{equation} 
\label{Eq2a}
[\hat{\mathcal{H}}] = [\hat{\mathcal{H}}_{\sigma}] + [\hat{\mathcal{H}}_F] + [\hat{\mathcal{H}}_{\sigma F}] + [\hat{\mathcal{H}}_{\sigma \sigma}].
\end{equation}
The  term $[\hat{\mathcal{H}}_{\sigma}]$ represents the operator of $\sigma$-polaron subsystem energy in the absence of interaction between $\sigma$-polarons  themselves and with EM-field. The second term is the Hamiltonian of  free EM-field, the third term describes the interaction of $\sigma$-polaron sybsystem with EM-field. The  term $[\hat{\mathcal{H}}_{\sigma\sigma}]$  characterizes intrachain  and interchain  polaron-polaron interaction.
Solution of the nonstationary hypercomplex Schr\"odinger equation for the state vector $[\left| {\Psi (t)} \right\rangle]$ with given Hamiltonian  in continuum limit   can be represented in the form of sum of $n$ solutions for $n$ chains,
where the solution for $q$-th chain $\tilde{\Phi}^l_q(x,t)$ is \cite{Part1}
\begin{equation}
\label{Eq3a}
\tilde{\Phi}^l_q(x,t) = \sum_{p = 0}^{n-1}\Phi^l_{qp}(x,t)[e_1]^p,
\end{equation}
in which the  matrix elements $\Phi^l_{qp}(x,t)$ are  determined by 
\begin{equation}
\begin{split}
\label{Eq4a}
&\Phi^l_{qp}(x,t) = \int\limits_{-\infty }^{\infty}\Theta^l_{q}(h,0) \exp{ \frac{-2\pi qpi}{n}} \exp{ihx}\times \\
&\exp{\{i\sum_{j = 0}^{n-1}\exp{\frac{2\pi qji}{n}(\vartheta_j(h) - g\sqrt{l-1}\kappa_j(h))}\}}dh,
\end{split}
\end{equation}
where $\Theta^l_{q}(h,0)$,$\vartheta_j(h)$, $\kappa_j(h)$ are determined by eigenvalues $\textbf{k}_\alpha \in C, \alpha = \overline{0, n-1}$ of $n$-numbers $\Phi^l(h,0)$, $\theta(h)$ and $\chi(h)$, 
$[e_1]^j$ is j-th power of the circulant matrix $[e_1]$
\begin{equation}
\label{Eq5a}
[e_1]=\left[\begin{array} {*{20}c} 0&1&0& ...&0  \\ 0&0&1& ...&0 \\ &...& \\ 0&0& ... &0&1\\1&0&...&0&0 \end{array}\right].
\end{equation}
From  equation (\ref{Eq3}) follows temporal dependence of the integral inversion for  jth chain in NT  and spectral dependence of the RS-signal amplitude, which explain the appearance of a number lines, which are additional to AFSWR lines in the spectra observed. 
The identification of the lines $1215{\pm}1$ $cm^{-1}$, $1779.5{\pm}1$ $cm^{-1}$,  $2022.3{\pm}0.5$ $cm^{-1}$ (and corresponding lines in boron ion modified sample) with Fourier-image of revival part of Rabi packet is confirmed by the following. It is well known, that  in the case of point absorbing centers Rabi frequency is linear function of the amplitude of oscillating EM-field.  Given dependency takes also place for the center of Rabi packet, that follows from the analysis of Fourier transform of temporal dependence  of the  integral inversion. Further, it is evident, that amplitude of laser wave, penetrating in IBMR, is lesser by excitation from unimplanted side in consequence of some absorption in the unimplanted volume of diamond crystal. We see, that really, the  values 1569 ${\pm}3$ $cm^{-1}$, 1757${\pm}5$ $cm^{-1}$ of two high frequency components are substantially less in given case, than $1779.5{\pm}1$ $cm^{-1}$ and  $2022.3{\pm}0.5$ $cm^{-1}$, observed by excitation from implanted side, Fig.2,  Fig.1. Relative frequency changes are 1,151 and 1.134 ${\pm}0.003$ for the pair [$2022.3{\pm}0.5$ $cm^{-1}$, 1757${\pm}5$] and [$1779.5{\pm}1$ $cm^{-1}$, 1569 ${\pm}3$] correspondingly and it is seen, that they are close to each other, at that  larger frequency undergoes larger change in correspondence with theoretical analysis.   The lines at 354.6,  977.1 (${\pm}1$ $cm^{-1}$ seem to be assigned with two AFESWR 
modes. The substantial decrease of relative intensity of 641.8 ${\pm}1$ $cm^{-1}$ mode in comparison with  $656.8{\pm}0.2$ $cm^{-1}$ mode testifies in favour of given assignment. It is seen, that AFESWR-splitting is rather large and it is comparable with the splitting between two polaron vibronic levels. It means, that linear AFESWR-theory  \cite{Yearchuck_PL},  which predicts   a set of equidistant  AFESWR-modes, arranged the left and the right of central mode, will be not applicable for given case. Really,  AFESWR-modes   are shifted on different distances 335.3, 287,2 $cm^{-1}$ from main AFESWR-mode. At the same time average value of AFESWR-splitting is 311.3$cm^{-1}$ and it is close to the value in 300 $cm^{-1}$  expected.   Therefore, we obtain the direct proof of assigment of the lines 656.8${\pm}0.2$ $cm^{-1}$ and 641.8 ${\pm}1$ $cm^{-1}$ with $\sigma$-polaron lattice.   

To explain  differences  by the change of the direction of the excitation wave propagation, we have to take into consideration the following. Any SWR-excitation in 1D-systems is strongly dependent on  the   directions of chain axis, vectors of intracrystalline external electric field, it is confirmed by ferromagnetic SWR study in carbynes, \cite{Ertchak_J_Physics_Condensed_Matter}. The axes in CZSNTs are not linear in the end of ion run and they are generatrixes  of the figure of onion-like shape, that provides for necessary geometry for  AFESWR-excitation of $\sigma$-polaron lattice. Moreover the appearance of very broad line seems to be indication on the excitation  of the Fr\"ohlich movement of $\sigma$-polaron lattice itself. The presence of Fr\"ohlich sliding of $\sigma$-polaron lattice allows to explain qualitatively the appearance of "hysteresis" relatively the direction of excitation wave propagation from energy law conservation position. Really, moving $\sigma$-polaron lattice possesses by kinetic energy. It means, that it is required lesser energy value, to excite the local polaron vibration mode in correspondence with observation. 

Therefore CZSNTs  represent themselves the example of the system, which strongly interact with EM-field. Experimental detection of Rabi wave packets confirms on the one hand the theory, developed in \cite{Part1} and it is also the first experimental confirmation of Rabi wave phenomenon, predicted in \cite{Slepyan_Yerchak}, on the other hand.  It means also, that semiclassical description of spectroscopic transitions in CZSNTs and in the systems like them cannot be appropriate, it substantially  raises the practical concernment of QED-theory.   The requirement of perfect onedimemensionality is substantial for applicability of SSH-model to CZSNTs, at the same time the Rabi wave phenomenon seems to be general and can be observed in 2D and 3D-systems. For instance, the first-order RS-peak in the vicinity of 1580 $cm^{-1}$ and so called second-order lines  around 2480, 2700 (2D band) , 3250 $cm^{-1}$ by $\lambda_{exc}$ = 488 nm were observed in graphene \cite{Calizo}, at that the intensity of 2D band exceeds substantially the intensty of the first-order line. Moreover theoretical dependence of 2D band position on excitation energy, presented  in \cite{Calizo} (Fig.5), is far from experimental dependence. At the same time the suggestion, that the second-order lines are spectral mapping  of revival part of Rabi wave packet is agreeing  with data in Fig.5 very well, since, if to choose instead energy the electrical component of oscillating excitation-field to be x-coordinate,  then we obtain linear experimental dependence in correspondence with expected for the center of Rabi packets. Similar explanation can be proposed for   transitions in heavily boron-doped
diamond in the region of 1400-2800 $cm^{-1}$ \cite{{Vlasov}}. Naturally  Rabi waves can also be observed in CVD-NTs.  

 Thus, QED-model for  multichain coupled qubit quasi-1D  system, proposed in \cite{Part1}, is confirmed by Raman scattering studies of carbon zigzag-shaped nanotubes, produced by HEIM of natural diamond single crystals. New quantum optics phenomenon - Rabi waves, predicted in \cite{Slepyan_Yerchak},  has been experimentally identified for the first time.  

\end{document}